\documentclass[12pt]{article}

\makeatletter

\@addtoreset{equation}{section} \makeatother
\newtheorem{theo}{Theorem}[section]
\newtheorem{rema}[theo]{Remark}

\newtheorem{propo}[theo]{Proposition}
\newtheorem{coro}[theo]{Corollary}

\newcommand{\be}{\begin{equation}}
\newcommand{\ee}{\end{equation}}

\font\ddpp=msbm10

\input{tcilatex}
\begin{document}

\title{\textbf{On the general structure of Ricci collineations for type B warped
spacetimes}}
\author{J.L. FLORES$^{1}$, Y. PARRA$^{2}$ and U. PERCOCO$^{3}$ \\
\\
{\small $^{1}$Departamento de Geometr\'{\i}a y Topolog\'{\i}a} \\
{\small Facultad de Ciencias, Universidad de Granada} \\
{\small Avenida Fuentenueva s/n, 18071 Granada, Spain} \\
{\small $^{2}$Laboratorio de F\'{\i}sica Te\'orica}\\
{\small Departamento de F\'{\i}sica, Facultad de Ciencias}\\
{\small Universidad de los Andes, M\'erida 5101, Venezuela}\\
{\small $^{3}$Centro de Astrof\'{\i}sica Te\'orica}\\
{\small Departamento de F\'{\i}sica, Facultad de Ciencias}\\
{\small Universidad de los Andes, M\'erida 5101, Venezuela}\\
{\small floresj@math.sunysb.edu, jparra@ula.ve, upercoco@ula.ve}}
\date{}
\maketitle

\textwidth=140mm \textheight=200mm \parindent=5mm

\begin{center}
{\small \textbf{Abstract}}
\end{center}

{\small A complete study of the structure of Ricci collineations for type B
warped spacetimes is carried out. This study can be used as a method to
obtain these symetries in such spacetimes. Special cases as $2+2$ reducible
spacetimes, and plane and spherical symmetric spacetimes are considered
specifically. } \newline


\newpage

\section{Introduction}

\label{sect1}

In the last years, symmetries in General Relativity have been studied in
depth because of their interest from both a mathematical and a physical
viewpoint. In fact, symmetries are important not only because of their
classical physical significance, but also because they simplify Einstein
equations and provide a classification of the spacetimes according to the
structure of the corresponding Lie algebra. They are described by vector
fields $X$ on the spacetime which satisfy a relation of the form: 
\[
\pounds _{X}\Phi =\Lambda , 
\]
where $\Phi $ is any of the quantities $g_{ab}$, $R_{ab}$, $R_{bcd}^{a}$,
etc, $\Lambda $ is a tensor with the same index symmetries as $\Phi $ and $%
\pounds $ represents the Lie derivative. Depending on $\Phi $ and $\Lambda $%
, there are different classes of symmetries (the relation between them was
studied in \cite{KatzinEtal69}). For example, if $\Phi =g_{ab}$ and $\Lambda
=\psi g_{ab}$, with $\psi$ a function, then $X$ is a \emph{Killing} vector
field if $\psi =0$, a \emph{homothetic} vector field if $\psi _{,a}=0$, a 
\emph{special conformal} vector field if $\psi _{;ab}=0$, and a \emph{%
conformal} vector field if $\psi $ is arbitrary. A symmetry will be called 
\emph{proper} if it does not belong to any of its subtypes, otherwise it
will be said \emph{improper}.

In this article we will concentrate on \emph{Ricci collineations}, that is,
the case when $\Phi =R_{ab}$ and $\Lambda =0$. These symmetries are
interesting because, among other things, they provide information about the
energy-momentum tensor via the Einstein equations (although Ricci
collineations are not usually matter collineations). In order to ensure that
Ricci collineations form a Lie algebra with the usual bracket operation, we
shall assume that they are smooth vector fields. Recall that this algebra
naturally contains all the special conformal vector fields (and thus, all
the homothetic and Killing vector fields). Regarding the Ricci tensor, we
shall consider (up to Section \ref{degenerate}) that it is non-degenerate,
i.e. rank 4; in particular, this ensures that the corresponding Lie algebra
is finite-dimensional, with maximal dimension being 10. Further information
on dimensionality and degenerate Ricci tensors can be found, for example, in
references \cite{CarotEtal94}, \cite{HallEtal95}.


In \cite{camci} the general form and classification of Ricci collineations
of Robertson-Walker spacetimes is provided in detail. Afterwards, in \cite
{metodo} the authors compute Ricci collineations of metrics $g_{ab}$ which
are conformal to $1+(n-1)$ decomposable metrics by using an interesting
technique. Roughly speaking, they construct the generic metric $G_{ab}$
defined from the symmetry group of $g_{ab}$. Then, proper Ricci
collineations are the Killing vector fields of $G_{ab}$ which are not
Killing vector fields of $g_{ab}$. This method provides the Ricci
collineations of Robertson-Walker spacetimes without any further
calculations.

A few years ago, the problem of determining all Ricci collineations of type
B warped spacetimes was considered in \cite{Carotetal}. This class of
spacetimes is important because its structure is satisfied by multiple
examples of interest in Physics as Schwarzschild, Robertson-Walker, etc.
Unfortunately, the conclusion obtained there cannot be considered the
solution of the problem, because it does not give \textit{all} Ricci
collineations of such spacetimes. In fact, two simple counterexamples to the
main result in \cite{Carotetal} were given in \cite{comtocar}. On the other
hand, the technique introduced in \cite{metodo} does not seem to be
applicable directly to these spacetimes. In conclusion, this problem remains
still open.

Our aim in this article is to describe in a general context the structure of
all Ricci collineations of type B warped spacetimes. In fact, after a study
of the equations which define these symmetries in such spacetimes, we
classify them according to their structure. This classification can be
considered a method to obtain all Ricci collineations. In particular, the
counterexamples given in \cite{comtocar} are clearly contained in our
results, Remark \ref{rema1} (1). This article is organized as follows.

After some preliminaries on type B warped spacetimes, in Section \ref{sect2}
we obtain two conclusions (Propositions \ref{t1} and \ref{t2}) on the
structure of Killing vector fields and Ricci collineations of such
spacetimes. In Section \ref{sect3}, these results are applied
systematically. Ricci collineations are classified according to having or
not mixed variables and, in each case, according to their vertical
component. As consequence, an exhaustive description of the structure of
these symmetries is obtained. In Section \ref{sect4}, $2+2$ reducible
spacetimes (Subsection \ref{subsect4.1}) and plane and spherical symmetric
spacetimes (Subsection \ref{subsect4.2}) are studied specifically. Finally,
the case when Ricci tensor is degenerate is briefly considered in Section 5.

\section{Preliminaries}

\label{sect2}

Let $(M_{1},g_{1})$ and $(M_{2},g_{2})$ be semi-Riemannian manifolds, and $%
\phi>0$ a smooth function on $M_{1}$. A \textit{warped product} with base $%
(M_{1},g_{1})$, fiber $(M_{2},g_{2})$ and warping function $\phi>0$ is the
product manifold $M=M_{1}\times M_{2}$ endowed with the metric tensor: 
\[
g^{\phi}=\pi_{1}^{*}g_{1}+(\phi\circ\pi_{1})^{2}\pi^{*}_{2}g_{2}\equiv
g_{1}+\phi^{2}g_{2}, 
\]
where $\pi_{1}$ and $\pi_{2}$ are the natural projections of $M_{1}\times
M_{2}$ onto $M_{1}$ and $M_{2}$, respectively. If, additionally, $%
(M,g^{\phi})$ is a connected time-oriented four-dimensional Lorentzian
manifold, then we say that $(M,g^{\phi})$ is a \textit{warped spacetime}. In
this case, a classification can be made according to the respective
dimensions of $M_{1}$ and $M_{2}$ (see \cite{CarotDaCosta93} and references
therein for a general discussion).

In this article we will concentrate on the study of \textit{type B warped
spacetimes}, that is, the case when $M_{1}$ and $M_{2}$ are both of
dimension $2$. In this case, and whenever we work locally, we can assume: 
\[
g^{\phi}=g_{AB}(x^{C})dx^{A}dx^{B}+\phi^{2}(x^{C})g_{\alpha\beta}(x^{%
\gamma})dx^{\alpha}dx^{\beta}\quad 
\begin{array}{l}
A,B,C=0,1 \\ 
\alpha,\beta,\gamma=2,3
\end{array}
\]
where $g_{AB}$ and $g_{\alpha\beta}$ are the components of $g_{1}$ and $%
g_{2} $ in certain charts $(U_{1}\subseteq M_{1},x^{0},x^{1})$, $%
(U_{2}\subseteq M_{2},x^{2},x^{3})$, respectively.

Let $X$ be a vector field on $M$ and consider its horizontal and vertical
components $X_{1}$, $X_{2}$; that is, 
\[
X_{1}(x^{A},x^{\alpha})=d\pi_{1}(X)(x^{A},x^{\alpha})\quad\quad
X_{2}(x^{A},x^{\alpha})=d\pi_{2}(X)(x^{A},x^{\alpha}). 
\]
Then, the Lie derivative of $g^{\phi}$ with respect to $X$ is: 
\begin{equation}  \label{liem1}
(\pounds_{X} g^{\phi})_{AB}=(\pounds _{X_1}g_{1})_{AB}\quad\quad\quad\quad
\end{equation}
\begin{equation}  \label{liem2}
\quad(\pounds_{X}g^{\phi})_{A\alpha}=g_{AC}X^C_{1,\alpha}+\phi^{2}g_{\alpha%
\beta}X^{\beta}_{2,A}
\end{equation}
\begin{equation}  \label{liem3}
\quad\quad\quad(\pounds_{X} g^{\phi})_{\alpha\beta}=\phi^{2}(\pounds
_{X_2}g_{2})_{\alpha\beta}+\phi^{2}_{,C}X_1^C g_{\alpha\beta}.
\end{equation}
In order to find the Killing vector fields of $(M,g^{\phi})$, (\ref{liem1}),
(\ref{liem2}) and (\ref{liem3}) must be set equal to zero. Condition (\ref
{liem1}) equal to zero is equivalent to: for every $p_{2}\in M_{2}$ the
restriction of $X_{1}$ to $M_{1}\times p_{2}$ is a Killing vector field
(perhaps zero) of $(M_{1}\times p_{2},g_{1})$. On the other hand, (\ref
{liem3}) equal to zero is equivalent to: for every $p_{1}\in M_{1}$ the
restriction of $X_{2}$ to $p_{1}\times M_{2}$ is a conformal vector field
(perhaps zero) of $(p_{1}\times M_{2},g_{2})$ with conformal factor 
\[
\psi=-\frac{1}{2}\frac{\phi^{2}_{,C}X^{C}_{1}}{\phi^{2}}. 
\]
These simple facts are summarized in the following way:

\begin{propo}
\label{t1} Let $(M,g^{\phi})$ be a type B warped spacetime with base $%
(M_{1},g_{1})$, fiber $(M_{2},g_{2})$ and warping function $\phi>0$. A
vector field $X\not{\equiv }0$ on $M$ is Killing of $(M,g^{\phi})$ if and
only if the following statements hold:

\begin{itemize}
\item[(i)]  for every $p_{1}\in M_{1}$, $X_{2}$ is a conformal vector field
(perhaps zero) of $(p_{1}\times M_{2},g_{2})$ with conformal factor $\psi$,

\item[(ii)]  for every $p_{2}\in M_{2}$, $X_{1}$ is a Killing vector field
(perhaps zero) of $(M_{1}\times p_{2},g_{1})$, which satisfies 
\begin{equation}  \label{star}
\phi _{,C}^{2}X_{1}^{C}=-2\psi \phi ^{2}
\end{equation}
and,

\item[(iii)]  components (\ref{liem2}) are equal to zero.
\end{itemize}
\end{propo}

A direct computation provides the following components $R_{ab}$ of the Ricci
tensor $\mathbf{R}$ of a type B warped spacetime: 
\begin{equation}  \label{appear}
\begin{array}{l}
R_{AB}=\frac 12R_1 g_{AB}-\frac 2\phi \phi _{A;B} \\ 
R_{A\alpha }=0 \\ 
R_{\alpha \beta }=\frac 12\left( R_2-\left( \phi ^2\right) _{;A}^A\right)
g_{\alpha \beta }\ \equiv \ Fg_{\alpha \beta},
\end{array}
\end{equation}
where, obviously, $F:=\frac 12\left( R_2-\left( \phi ^2\right)
_{;A}^A\right) $, $R_1$ and $R_2$ are the scalar curvatures of $g_1$ and $%
g_2 $, respectively, and the semi-colon indicates the covariant derivative 
\textit{with respect to} $g^{\phi}$.

\begin{rema}
\label{r1} \emph{Although the terms $\phi _{A;B}$ and $(\phi ^{2})_{;A}^{A}$
in (\ref{appear}) include covariant derivatives with respect to all the
metric $g^{\phi }$, a direct computation shows that they are independent of
the variables $x^{\gamma}$ of $M_{2}$. In fact: 
\[
\begin{array}{l}
\phi_{A;B}=\phi_{,AB}-\frac{g^{CD}}{2}(g_{DB,A}+g_{DA,B}-g_{AB,D})\phi_{,C}
\\ 
(\phi^{2})^{A}_{;A}=g^{AB}\phi^{2}_{,AB}-\frac{g^{AB}g^{CD}}{2}%
(g_{DB,A}+g_{DA,B}-g_{AB,D})\phi^{2}_{,C}.
\end{array}
\]
}
\end{rema}

The Lie derivative of $\mathbf{R}$ with respect to $X$ is: 
\begin{equation}  \label{lie1}
(\pounds_{X} \mathbf{R})_{AB}=R_{AB,C}X_1^C + R_{AC}X^C_{1,B} +
R_{CB}X^C_{1,A}\quad\;\;
\end{equation}
\begin{equation}  \label{lie2}
(\pounds_{X} \mathbf{R})_{A\alpha}=R_{AC}X^C_{1,\alpha}+R_{\alpha\beta}X^{%
\beta}_{2,A}\;\;\quad\quad\quad\quad\quad\quad\quad
\end{equation}
\begin{equation}  \label{lie3}
(\pounds_{X} \mathbf{R})_{\alpha\beta}=F(\pounds
_{X_2}g_{2})_{\alpha\beta}+F_{,C}X_1^C g_{\alpha\beta}+F_{,\gamma
}X_2^\gamma g_{\alpha\beta}.
\end{equation}
In the following, our aim will be to find the Ricci collineations of $%
(M,g^{\phi})$; that is, the vector fields $X\not{\equiv }0$ on $M$ such that
(\ref{lie1}), (\ref{lie2}) and (\ref{lie3}) are equal to zero. 

As commented in the Introduction, we will assume that $\mathbf{R}$ is
non-degenerate. Therefore, $F\neq 0$ everywhere. Moreover, from (\ref{appear}%
) and Remark \ref{r1}, $R_{AB}$ can be seen as the components of a metric
tensor $g_{R}$ defined on $M_{1}$. Then, reasoning as in Proposition \ref{t1}%
, condition (\ref{lie1}) equal to zero is equivalent to: for every $p_{2}\in
M_{2}$ the restriction of $X_{1}$ to $M_{1}\times p_{2}$ is a Killing vector
field (perhaps zero) of $(M_{1}\times p_{2},g_{R})$. On the other hand, (\ref
{lie3}) equal to zero is equivalent to: for every $p_{1}\in M_{1}$ the
restriction of $X_{2}$ to $p_{1}\times M_{2}$ is a conformal vector field
(perhaps zero) of $(p_{1}\times M_{2},g_{2})$ with conformal factor 
\begin{equation}
\psi =-\frac{1}{2}{\frac{F_{,C}X_{1}^{C}+F_{,\gamma }X_{2}^{\gamma }}{F}}.
\label{conformalfactor}
\end{equation}
Equation (\ref{conformalfactor}) can be simplified by using the classical
expression of the Lie derivative of the Ricci $\mathbf{R}_{h}$ of a
semi-Riemannian manifold $(N,h)$ with respect to a conformal vector field $Y$
of conformal factor $\xi$ (see \cite{Hall90}); that is, 
\begin{equation}  \label{classic}
(\pounds _{Y} \mathbf{R}_{h})_{ab}=-(n-2)\xi _{|ab}-(\Delta _{h}\xi )h_{ab},
\end{equation}
where $n=\emph{dim}\,N$ and $\Delta _{h}\xi =\xi _{|cd}h^{cd}$ is the
Laplacian of $\xi $ with respect to $h$ (obviously, the stroke denotes the
covariant derivative with respect to $h$). In fact, assume that $X_{2}$ is a
conformal vector field of $(p_{1}\times M_{2},g_{2})$ with conformal factor $%
\psi$. Then, from (\ref{classic}) we obtain: 
\[
(\pounds _{X_{2}}\mathbf{R}_{g_{2}}) _{\alpha \beta}=-(\Delta _{g_{2}}\psi
)g_{\alpha \beta }. 
\]
But, obviously, 
\[
\pounds_{X_{2}}(\mathbf{R}_{g_{2}})_{\alpha\beta}=\pounds _{X_{2}}\left( 
\frac{1}{2}R_{2}g_{2}\right) _{\alpha \beta }=\frac{1}{2}(R_{2,\gamma
}X_{2}^{\gamma }+2\psi R_{2})g_{\alpha \beta }; 
\]
thus 
\begin{equation}
R_{2,\gamma }X_{2}^{\gamma }+2\psi R_{2}=-2\Delta _{g_{2}}\psi .
\label{masutil}
\end{equation}
On the other hand, by replacing in (\ref{conformalfactor}) the expression of 
$F$ we have: 
\begin{equation}  \label{cuenta}
2\psi(R_{2}-(\phi^{2})^{A}_{;A})=(\phi^{2})^{A}_{;A,C}X_{1}^{C}-R_{2,%
\gamma}X_{2}^{\gamma}.
\end{equation}
Therefore, from (\ref{masutil}) and (\ref{cuenta}) we obtain: 
\begin{equation}
(\phi ^{2})_{;A,C}^{A}X_{1}^{C}=-2\psi (\phi ^{2})_{;A}^{A}-2\Delta
_{g_{2}}\psi .  \label{massutil}
\end{equation}
These facts are summarized in the following result:

\begin{propo}
\label{t2} Let $(M,g^{\phi})$ be a type B warped spacetime with base $%
(M_{1},g_{1})$, fiber $(M_{2},g_{2})$ and warping function $\phi>0$. A
vector field $X\not{\equiv }0$ on $M$ is a Ricci collineation of $%
(M,g^{\phi})$ if and only if the following statements hold:

\begin{itemize}
\item[(i)]  for every $p_{1}\in M_{1}$, $X_{2}$ is a conformal vector field
(perhaps zero) of $(p_{1}\times M_{2},g_{2})$ with conformal factor $\psi$,

\item[(ii)]  for every $p_{2}\in M_{2}$, $X_{1}$ is a Killing vector field
(perhaps zero) of $(M_{1}\times p_{2},g_{R})$, which satisfies (\ref
{massutil}), and

\item[(iii)]  components (\ref{lie2}) are equal to zero.
\end{itemize}
\end{propo}

In the next section, Propositions \ref{t1} and \ref{t2} will be exploited in
order to describe the general structure of Ricci collineations of $%
(M,g^{\phi })$.

\section{Ricci collineations of type B warped spacetimes}

\label{sect3}

For simplicity, firstly we will classify these symmetries in two families.
In the first family, we will include Ricci collineations whose variables are
not mixed, that is, when the corresponding vector field $X$ can be written
as 
\[
X(x^{A},x^{\alpha})=X_{1}(x^{A})+X_{2}(x^{\alpha}). 
\]
Our study is completed by including in a second family Ricci collineations
such that either $\partial X_{1}/\partial x^{\alpha}\neq 0$ or $\partial
X_{2}/\partial x^{A}\neq 0$.

\vspace{5mm}

\noindent \textbf{FAMILY 1.} Ricci collineations with non-mixed variables.

\vspace{2mm}

Notice that, in this case, statements $(iii)$ in Propositions \ref{t1} and 
\ref{t2} always hold. On the other hand, from Proposition \ref{t2}, we can
distinguish four types in this family attending to the vertical component $%
X_{2}$ of $X$.

\vspace{3mm}

\noindent \textbf{TYPE 1.1:} $X_{2}$ is a Killing vector field (perhaps
zero) of $(M_{2},g_{2})$.

\vspace{2mm}

>From Proposition \ref{t2} $(ii)$, $X\not{\equiv }0$ will be a Ricci
collineation if, additionally, $X_{1}$ is a Killing vector field (perhaps
zero) of $(M_{1},g_{R})$ with $(\phi ^{2})_{;A,C}^{A}X_{1}^{C}=0$.
Therefore, Ricci collineations $X\not{\equiv }0$ of type 1.1 are: 
\[
X=X_{1}+X_{2}=\sum_{i=1}^{k_{R}}a_{i}^{1}X_{1}^{i}+%
\sum_{j=1}^{k_{2}}a_{j}^{2}X_{2}^{j} 
\]
where

\begin{itemize}
\item[(i)]  $\{X^{i}_{1}\}_{i=1}^{k_{R}}$ is the algebra of Killing vector
fields of $(M_{1},g_{R})$,

\item[(ii)]  $\{X_{2}^{j}\}_{j=1}^{k_{2}}$ is the algebra of Killing vector
fields of $(M_{2},g_{2})$, and

\item[(iii)]  coefficients $\{a^{1}_{i}\}_{i=1}^{k_{R}}$ satisfy 
\begin{equation}  \label{rr1}
\sum_{i=1}^{k_{R}}a_{i}^{1}(\phi^{2})^{A}_{;A,C}X_{1}^{i\,C}=0.
\end{equation}
\end{itemize}

Additionally, from Proposition \ref{t1} $X$ is not a Killing vector field of 
$(M,g^{\phi})$ if,

\begin{itemize}
\item[(iv)]  either $\phi^{2}_{,C}X_{1}^{C}\neq 0$ or $X_{1}$ is not a
Killing vector field of $(M_{1},g_{1})$ (in particular, $X_{1}\not{\equiv }0$%
).
\end{itemize}

\begin{rema}
\emph{As $\emph{dim}\,M_{i}=2$, $i=1,2$, necessarily $k_{R},k_{2}=0,1,3$.
But, from (iv), $(M,g^{\phi})$ admits proper Ricci collineations of type 1.1
only if $k_{R}=1,3$. Therefore, in this case, if the curvature of $%
(M_{1},g_{R})$ is not constant, necessarily $k_{R}=1$, and thus, equation (%
\ref{rr1}) reduces to $(\phi ^{2})_{;A,C}^{A}X_{1}^{1\,C}=0$. }
\end{rema}

\vspace{3mm}

\noindent \textbf{TYPE 1.2:} $X_{2}$ is a proper homothetic vector field of $%
(M_{2},g_{2})$.

\vspace{2mm}

Obviously, this type of collineations only exists if the curvature of $%
(M_{2},g_{2})$ is not a constant different from zero. In this case, $X$ will
be a Ricci collineation if, additionally, $X_{1}$ is a Killing vector field
(perhaps zero) of $(M_{1},g_{R})$ with 
\begin{equation}
(\phi ^{2})_{;A,C}^{A}X_{1}^{C}=-2\lambda (\phi ^{2})_{;A}^{A},  \label{rere}
\end{equation}
where $\lambda\neq 0$ is the homothetic factor of $X_{2}$. From (\ref{rere}%
), recall that if $X_{1}=0$ then, necessarily $(\phi^{2})^{A}_{;A}=0$.

In conclusion, Ricci collineations $X\not{\equiv }0$ of type 1.2 are: 
\[
X=X_{1}+X_{2}=\sum_{i=1}^{k_{R}}a_{i}^{1}X_{1}^{i}+%
\sum_{j=1}^{k_{2}}a_{j}^{2}X_{2}^{j}+\lambda Y, 
\]
where,

\begin{itemize}
\item[(i)]  as before, $\{X_{1}^{i}\}_{i=1}^{k_{R}}$, $\{X_{2}^{j}%
\}_{j=1}^{k_{2}}$ are the algebras of Killing vector fields of $%
(M_{1},g_{R}) $, $(M_{2},g_{2})$, respectively,

\item[(ii)]  $Y$ is the homothetic vector field of $(M_{2},g_{2})$ with
homothetic factor $1$, and

\item[(iii)]  coefficients $\{a_{i}^{1}\}_{i=1}^{k_{R}}$ and $\lambda\neq 0$
satisfy 
\[
\sum_{i=1}^{k_{R}}a_{i}^{1}(\phi^{2})^{A}_{;A,C}X_{1}^{i\,C}=-2\lambda(%
\phi^{2})^{A}_{;A}. 
\]
\end{itemize}

Additionally, $X$ is not a Killing vector field of $(M,g^{\phi})$ if,

\begin{itemize}
\item[(iv)]  either $\phi^{2}_{,C}X_{1}^{C}\neq -2\lambda\phi^{2}$ or $X_{1}$
is not a Killing vector field of $(M_{1},g_{1})$.
\end{itemize}

\vspace{3mm}

\noindent \textbf{TYPE 1.3:} $X_{2}$ is a proper special conformal vector
field of $(M_{2},g_{2})$.

\vspace{2mm}

This type of collineations only exists if $(\phi^{2})^{A}_{;A}=0$. In fact,
now the conformal factor $\psi$ associated to $X_{2}$ is a non-constant
function of $x^{\alpha}$ with $\Delta_{g_{2}}\psi=0$. Therefore, if we
assume that (\ref{massutil}) holds, and derive it with respect to $%
x^{\gamma} $, we deduce that $(\phi^{2})^{A}_{;A}=0$.

Under this restriction, all Ricci collineations $X\not{\equiv }0$ of type
1.3 are given by: 
\[
X=X_{1}+X_{2}=\sum_{i=1}^{k_{R}}a_{i}^{1}X_{1}^{i}+%
\sum_{j=1}^{s_{2}}a_{j}^{2}X_{2}^{j}, 
\]
where,

\begin{itemize}
\item[(i)]  $\{X_{1}^{i}\}_{i=1}^{k_{R}}$ is the algebra of Killing vector
fields of $(M_{1},g_{R})$,

\item[(ii)]  $\{X_{2}^{j}\}_{j=1}^{s_{2}}$ is the algebra of special
conformal vector fields of $(M_{2},g_{2})$ and,

\item[(iii)]  some of the coefficients $\{a_{j}^{2}\}_{j=h_{2}+1}^{s_{2}}$
are different from zero, being $h_{2}$ the dimension of the algebra of
homothetic vector fields of $(M_{2},g_{2})$.
\end{itemize}

Moreover, these collineations are not Killing vector fields of $(M,g^{\phi})$
because they do not satisfy (\ref{star}).

\begin{rema}
\emph{Notice that condition $(\phi^{2})^{A}_{;A}=0$ implies that proper
homothetic and special conformal vector fields of $(M_{2},g_{2})$ are also
Ricci collineations of $(M,g^{\phi})$ of types 1.2 and 1.3, respectively.
Moreover, they are not Killing vector fields of $(M,g^{\phi})$ (since they
do not satisfy (\ref{star})).}
\end{rema}

\vspace{3mm}

\noindent \textbf{TYPE 1.4:} $X_{2}$ is a proper conformal vector field of $%
(M_{2},g_{2})$.

\vspace{2mm}

This type of collineations only exists if $(\phi^{2})^{A}_{;A}$ remains
constant wherever $\psi$ is not constant. In fact, if we assume that (\ref
{massutil}) holds, and derive it with respect to $x^{\gamma}$, we deduce
that 
\begin{equation}  \label{r4}
\Delta_{g_{2}} \psi=-(\phi^{2})^{A}_{;A}\cdot \psi=-\hbox{const}\cdot \psi
\end{equation}
on such a domain.

In conclusion, Ricci collineations $X\not{\equiv }0$ of type 1.4 satisfy the
expression: 
\[
X=X_{1}+X_{2}=\sum_{i=1}^{k_{R}}a_{i}^{1}X_{1}^{i}+%
\sum_{j=1}^{c_{2}}a_{j}^{2}X_{2}^{j}, 
\]
where

\begin{itemize}
\item[(i)]  $\{X_{1}^{i}\}_{i=1}^{k_{R}}$ is the algebra of Killing vector
fields of $(M_{1},g_{R})$,

\item[(ii)]  $\{X_{2}^{j}\}_{j=1}^{c_{2}}$ is the conformal algebra of $%
(M_{2},g_{2})$ and,

\item[(iii)]  coefficients $\{a_{j}^{2}\}_{j=k_{2}+1}^{c_{2}}$ are such that
the conformal factor of $X_{2}$ 
\[
\psi =\sum_{j=k_{2}+1}^{c_{2}}a_{j}^{2}\psi _{2}^{j} 
\]
satisfies (\ref{massutil}) (in particular, satisfies (\ref{r4}) wherever $%
\psi$ is not constant), where $\{\psi_{2}^{j}\}_{j=k_{2}+1}^{c_{2}}$ are the
corresponding conformal factors of $\{X_{2}^{j}\}_{j=k_{2}+1}^{c_{2}}$, and
some of the coefficients $\{a_{j}^{2}\}_{j=s_{2}+1}^{c_{2}}$ must be
different from zero.
\end{itemize}

Again, these collineations are not Killing vector fields of $(M,g^{\phi})$
because they do not satisfy (\ref{star}).

\vspace{5mm}

\noindent \textbf{FAMILY 2.} Ricci collineations with mixed variables.

\vspace{2mm}

In this family, the dependence of $X_{1}$ and $X_{2}$ is not restricted to $%
x^{A}$ and $x^{\alpha}$, respectively. Therefore, $(iii)$ in Propositions 
\ref{t1} and \ref{t2} must be also taken into account in order to find these
symmetries. Summarizing, Ricci collineations $X\not{\equiv }0$ are given now
by: 
\[
X=X_{1}+X_{2}=\sum_{i=1}^{k_{R}}a_{i}^{1}(x^{\alpha})X_{1}^{i}+%
\sum_{j=1}^{c_{2}}a_{j}^{2}(x^{A})X_{2}^{j} 
\]
where

\begin{itemize}
\item[(i)]  $\{X_{1}^{i}\}_{i=1}^{k_{R}}$ is the algebra of Killing vector
fields of $(M_{1},g_{R})$,

\item[(ii)]  $\{X_{2}^{j}\}_{j=1}^{c_{2}}$ is the conformal algebra of $%
(M_{2},g_{2})$, and

\item[(iii)]  functions $\{a_{i}^{1}(x^{\alpha})\}_{i=1}^{k_{R}}$, $%
\{a_{j}^{2}(x^{A})\}_{j=1}^{c_{2}}$ satisfy
\end{itemize}

\begin{equation}  \label{d4}
\sum_{i=1}^{k_{R}}a_{i}^{1}(x^{\alpha})(\phi^{2})^{A}_{;A,C}X_{1}^{i\,C}=-2%
\left(\sum_{j=k_{2}+1}^{c_{2}}a_{j}^{2}(x^{A})\psi_{2}^{j}\right)(%
\phi^{2})^{A}_{;A}-2\sum_{j=s_{2}+1}^{c_{2}}a^{2}_{j}(x^{A})\Delta_{g_{2}}%
\psi^{j}_{2},
\end{equation}
\begin{equation}  \label{44}
\sum_{i=1}^{k_{R}}\frac{da^{1}_{i}(x^{\gamma})}{dx^{\alpha}}%
R_{AC}X_{1}^{iC}+\sum_{j=1}^{c_{2}}\frac{da^{2}_{j}(x^{C})}{dx^{A}}%
R_{\alpha\beta}X_{2}^{j\beta}=0,\quad A=0,1,\;\alpha=2,3
\end{equation}
(the indexes $k_{2}$, $s_{2}$ are again the dimensions of the homothetic and
special conformal algebras).

\noindent Additionally, $X$ is not a Killing vector field of $(M,g^{\phi})$
if

\begin{itemize}
\item[(iv)]  any of the statements of Proposition \ref{t1} do not hold.
\end{itemize}

Analogously to Family 1, we classify these collineations in four types.

\vspace{3mm}

\noindent \textbf{TYPE 2.1:} For every $p_{1}\in M_{1}$, $X_{2}$ is a
Killing vector field (perhaps zero) of $(p_{1}\times M_{2},g_{2})$.

\vspace{2mm}

In this case the only functions which can be different from zero are $%
\{a_{i}^{1}(x^{\alpha })\}_{i=1}^{k_{R}}$, $\{a_{j}^{2}(x^{A})%
\}_{j=1}^{k_{2}}$. As a consequence, equation (\ref{d4}) reduces to 
\[
\sum_{i=1}^{k_{R}}a_{i}^{1}(x^{\alpha })(\phi
^{2})_{;A,C}^{A}X_{1}^{i\,C}=0. 
\]

\vspace{3mm}

\noindent \textbf{TYPE 2.2:} For every $p_{1}\in M_{1}$, $X_{2}$ is a
homothetic vector field (perhaps zero) of $(p_{1}\times M_{2},g_{2})$ which
is not always Killing.

\vspace{2mm}

This type of collineations only exists if the curvature of $(M_{2},g_{2})$
is not a constant different from zero. In this case, only the functions $%
\{a_{i}^{ 1}(x^{\alpha})\}_{i=1}^{k_{R}}$, $\{a_{j}^{
2}(x^{A})\}_{j=1}^{k_{2}}$, $a_{k_{2}+1}^{2}(x^{A})\equiv\lambda(x^{A})$ can
be different from zero, and they must satisfy: 
\[
\sum_{i=1}^{k_{R}}a_{i}^{1}(x^{\alpha})(\phi^{2})^{A}_{;A,C}X_{1}^{i\,C}=-2%
\lambda(x^{A})(\phi^{2})^{A}_{;A}, 
\]
where we have assumed the homothetic factor $\psi^{k_{2}+1}_{2}$ normalized
to $1$.

\vspace{3mm}

\noindent \textbf{TYPE 2.3:} For every $p_{1}\in M_{1}$, $X_{2}$ is a
special conformal vector field (perhaps zero) of $(p_{1}\times M_{2},g_{2})$
which is not always homothetic.

In this case, only the functions $\{a_{i}^{1}(x^{\alpha })\}_{i=1}^{k_{R}}$, 
$\{a_{j}^{2}(x^{A})\}_{j=1}^{s_{2}}$ can be different from zero. As a
consequence, equation (\ref{d4}) reduces now to: 
\[
\sum_{i=1}^{k_{R}}a_{i}^{1}(x^{\alpha })(\phi
^{2})_{;A,C}^{A}X_{1}^{i\,C}=-2\left(
\sum_{j=k_{2}+1}^{s_{2}}a_{j}^{2}(x^{A})\psi _{2}^{j}\right) (\phi
^{2})_{;A}^{A}. 
\]

\vspace{3mm}

\noindent \textbf{TYPE 2.4:} For every $p_{1}\in M_{1}$, $X_{2}$ is a
conformal vector field (perhaps zero) of $(p_{1}\times M_{2},g_{2})$ which
is not always special. In general, we cannot simplify the structure of these
Ricci collineations.

\vspace{3mm}

In the following section a brief application of our study to some examples
of type B warped spacetimes is carried out. Without any further
calculations, we obtain an interesting information about the particular
structure of their symmetries.

\section{Examples}

\label{sect4}

In this section we will apply our point of view to the following families of
type B warped spacetimes: $2+2$ reducible spacetimes, and plane and
spherical symmetric spacetimes.

\subsection{$2+2$ reducible spacetimes}

\label{subsect4.1}

In this case the product manifold $M=M_{1}\times M_{2}$ is endowed with the
metric tensor: 
\[
g=\pi_{1}^{*}g_{1}+\pi_{2}^{*}g_{2}\equiv g_{1}+g_{2}. 
\]
Therefore, these spacetimes are type B warped spacetimes with $\phi^{2}=1$
and, thus, we can apply our previous study. Firstly, take into account that
now $g_{R}=1/2 R_{1}g_{1}=\mathbf{R}_{g_{1}}$. Thus, Killing vector fields $%
X_{1}$ of $(M_{1},g_{R})$ are just the conformal vector fields of $%
(M_{1},g_{1})$ with conformal factors satisfying $\Delta_{g_{1}}\psi_{1}=0$
(recall (\ref{classic})). Therefore, if we apply Proposition \ref{t2} to
these spacetimes, we obtain the following consequences:

\begin{coro}
\label{c1} Ricci collineations $X\not{\equiv }0$, with non-mixed variables,
of a $2+2$ reducible spacetime $(M,g)$ are the vector fields $X=X_{1}+X_{2}$
such that, $X_{l}$ are conformal vector fields of $(M_{l},g_{l})$ with
conformal factors $\psi_{l}$ satisfying 
\[
\Delta_{g_{l}}\psi_{l}=0, \quad l=1,2. 
\]
\end{coro}

\begin{coro}
\label{c2} Ricci collineations $X\not{\equiv }0$, with mixed variables, of a 
$2+2$ reducible spacetime $(M,g)$ are the vector fields 
\[
X=X_{1}+X_{2}=\sum_{i=1}^{c_{1}}a_{i}^{1}(x^{\alpha})X_{1}^{i}+%
\sum_{j=1}^{c_{2}}a_{j}^{2}(x^{A})X_{2}^{j}, 
\]
where $\{X_{1}^{i}\}_{i=1}^{c_{1}}$, $\{X_{2}^{j}\}_{j=1}^{c_{2}}$ are the
conformal algebras of $(M_{1},g_{1})$, $(M_{2},g_{2})$, respectively, and
functions $\{a_{i}^{1}(x^{\alpha})\}_{i=1}^{c_{1}}$, $\{a
_{j}^{2}(x^{A})\}_{j=1}^{c_{2}}$ satisfy 
\[
\begin{array}{l}
\sum_{i=s_{1}+1}^{c_{1}}a_{i}^{1}(x^{\alpha})\Delta_{g_{1}}\psi^{i}_{1}=%
\sum_{j=s_{2}+1}^{c_{2}}a_{j}^{2}(x^{A})\Delta_{g_{2}}\psi^{j}_{2}=0, \\ 
\sum_{i=1}^{c_{1}}\frac{da_{i}^{1}(x^{\gamma})}{dx^{\alpha}}%
R_{AC}X_{1}^{iC}+\sum_{j=1}^{c_{2}}\frac{da_{j}^{2}(x^{C})}{dx^{A}}%
R_{\alpha\beta}X_{2}^{j\beta}=0, \quad A=0,1,\;\alpha=2,3,
\end{array}
\]
being $\{\psi_{1}^{i}\}_{i=1}^{c_{1}}$, $\{\psi_{2}^{j}\}_{j=1}^{c_{2}}$ the
corresponding conformal factors.
\end{coro}

\begin{rema}
\label{rema1} \emph{(1) Counterexamples given in \cite{comtocar} are clearly
contained in these results. In fact, $M=\hbox{\ddpp R}^{2}\times 
\hbox{\ddpp
R}^{2}$ endowed with 
\[
g=e^{t^{2}/2}(-dt^{2}+dx^{2})+e^{-y^{2}/2}(dy^{2}+dz^{2}) 
\]
is a $2+2$ reducible spacetime and, both, $X=\partial_{t}+\partial_{y}$, $%
Y=z\partial_{t}+\partial_{y}+t\partial_{z}$ are Ricci collineations which
satisfy hypotheses of Corollaries \ref{c1} and \ref{c2}, respectively. }

\emph{\noindent (2) From equations (\ref{appear}), it is clear that both
corollaries also hold for spacetimes not necessarily $2+2$ reducible, but
satisfying $\phi_{A;B}=(\phi^{2})^{A}_{;A}\equiv 0$.}
\end{rema}

\subsection{Plane and spherical symmetric spacetimes}

\label{subsect4.2}

Consider now the family of spacetimes $M=\hbox{\ddpp R}^{2}\times M_{2}$
endowed with the metric tensor: 
\[
g^{\phi }=-e^{2v}dt^{2}+e^{2w}dx^{2}+\phi ^{2}g_{2} 
\]
where $v,w$ and $\phi $ are each functions of $t$ and $x$, and 
\[
(M_{2},g_{2})=\left\{ 
\begin{array}{l}
\hbox{\ddpp R}^{2} \\ 
\hbox{\ddpp S}^{2}
\end{array}
\right. 
\]
endowed with their corresponding usual metrics (if we also include the
hyperbolic space, $g^{\phi}$ can be characterized by admitting a group $%
G_{3} $ acting multiply-transitively on spacelike orbits $V_{2}$, see \cite
{KramerEtal80}). To avoid the vanishing of $F$, and thus, the degeneracy of
Ricci tensor $\mathbf{R}$ (recall (\ref{appear})), we will also assume 
\begin{equation}
(\phi ^{2})_{;A}^{A}\neq R_{2}\quad \quad \hbox{for all}\;(t,x)\in %
\hbox{\ddpp R}^{2}.  \label{degeneration}
\end{equation}


>From Section \ref{sect3}, the component $X_{1}$ of a Ricci collineation $X%
\not{\equiv }0$ of these spacetimes is different from zero only if $%
(M_{1},g_{R})$ admits some Killing vector fields. In this case, the
dimension $k_{R}$ of the corresponding algebra must be 3 or 1, depending on
if $v$, $w$ and $\phi$ makes $(\hbox{\ddpp R}^{2},g_{R})$ being maximally
symmetric or not. To study the structure of $X_{2}$, we must consider the
two cases separately.

\vspace{3mm}

\textit{4.2.1. Plane symmetry:}

\vspace{2mm}

The conformal algebra of the plane $(\hbox{\ddpp R}^{2},dy^{2}+dz^{2})$ is
the (infinite-dimensional) \textit{Virasoro algebra}, which has the
following special conformal vector fields: 
\[
\begin{array}{ll}
X_{2}^{1}=\partial _{y} & \psi _{2}^{1}=0 \\ 
X_{2}^{2}=\partial _{z} & \psi _{2}^{2}=0 \\ 
X_{2}^{3}=z\partial _{y}-y\partial _{z} & \psi _{2}^{3}=0 \\ 
X_{2}^{4}=y\partial _{y}+z\partial _{z} & \psi _{2}^{4}=1 \\ 
X_{2}^{5}=(y^{2}-z^{2})\partial _{y}+2yz\partial _{z} & \psi _{2}^{5}=2y \\ 
X_{2}^{6}=2yz\partial _{y}+(z^{2}-y^{2})\partial _{z} & \psi _{2}^{6}=2z.
\end{array}
\]
Therefore, $k_{2}=3$, $h_{2}=4$, $s_{2}=6$ (and $c_{2}=\infty$). In this
case, we can establish the following:

\begin{itemize}
\item[(i)]  The vertical component $X_{2}$ of a Ricci collineation $X\not%
{\equiv }0$ of type 1.1 is a linear combination of the $k_{2}=3$ Killing
vector fields of the plane. On the other hand, the horizontal component $%
X_{1}$ satisfies the equation in $k_{R}$ variables (\ref{rr1}).

\item[(ii)]  The component $X_{2}$ of a Ricci collineation of type 1.2 is a
linear combination of the $h_{2}=4$ homothetic vector fields of the plane.
On the other hand, the horizontal component $X_{1}$ satisfies the equation (%
\ref{rere}). If $k_{R}=0$, there are not Ricci collineations of this type
since, in this case, (\ref{rere}) reduces to $(\phi ^{2})_{;A}^{A}=0$, which
contradicts (\ref{degeneration}).

\item[(iii)]  There are not Ricci collineations of type 1.3. Moreover, there
are collineations of type 1.4 only if $(\phi^{2})^{A}_{;A}=$const$\neq 0$.

\item[(iv)]  Ricci collineations in Family 2 must satisfy equations (\ref{d4}%
) and (\ref{44}), which are in general complicated. If $k_{R}=0$ and we
consider collineations of type 2.3, these equations reduce to: 
\[
a_{4}^{2}+2y\,a_{5}^{2}+2z\,a_{6}^{2}=0 
\]
and 
\[
\begin{array}{l}
a_{1,t}^{2}+z\,a_{3,t}^{2}+y\,a_{4,t}^{2}+(y^{2}-z^{2})a_{5,t}^{2}+2yz%
\,a_{6,t}^{2}=0 \\ 
a_{2,t}^{2}-y\,a_{3,t}^{2}+z\,a_{4,t}^{2}+2yz%
\,a_{5,t}^{2}+(z^{2}-y^{2})a_{6,t}^{2}=0 \\ 
a_{1,x}^{2}+z\,a_{3,x}^{2}+y\,a_{4,x}^{2}+(y^{2}-z^{2})a_{5,x}^{2}+2yz%
\,a_{6,x}^{2}=0 \\ 
a_{2,x}^{2}-y\,a_{3,x}^{2}+z\,a_{4,x}^{2}+2yz%
\,a_{5,x}^{2}+(z^{2}-y^{2})a_{6,x}^{2}=0.
\end{array}
\]
\end{itemize}

\vspace{3mm}

\textit{4.2.2. Spherical symmetry:}

\vspace{2mm}

In this case, the second space $(M_{2},g_{2})$ is the unitary bidimensional
sphere $\hbox{\ddpp S}^{2}$. The \emph{local} conformal algebra of $%
\hbox{\ddpp S}^{2}$, like that of the plane, is the Virasoro algebra. In
order to single out a finite-dimensional subalgebra from it, we will impose
that conformal vectors must be \emph{globally} defined on $\hbox{\ddpp S}%
^{2} $. Then, a well-known computation shows that the only \emph{global}
conformal vector fields of $\hbox{\ddpp S}^{2}$ expressed in spherical
coordinates are: 
\[
\begin{array}{ll}
X_{2}^{1}=\cos \varphi \partial _{\theta }-\sin \varphi \cot \theta \partial
_{\varphi } & \psi _{2}^{1}=0 \\ 
X_{2}^{2}=\sin \varphi \partial _{\theta }+\cos \varphi \cot \theta \partial
_{\varphi } & \psi _{2}^{2}=0 \\ 
X_{2}^{3}=\partial _{\varphi } & \psi _{2}^{3}=0 \\ 
X_{2}^{4}=\sin \theta \partial _{\theta } & \psi _{2}^{4}=\cos \theta \\ 
X_{2}^{5}=\cos \theta \cos \varphi \partial _{\theta }-\frac{\sin \varphi }{%
\sin \theta }\partial _{\varphi } & \psi _{2}^{5}=-\sin \theta \cos \varphi
\\ 
X_{2}^{6}=\cos \theta \sin \varphi \partial _{\theta }-\frac{\cos \varphi }{%
\sin \theta }\partial _{\varphi } & \psi _{2}^{6}=-\sin \theta \sin \varphi .
\end{array}
\]
(Nevertheless, recall that other conformal vectors---necessarily \emph{%
locally} defined---can appear as vertical component of a Ricci collineation
of a spherically symmetric spacetime.)

In conclusion, $k_{2}=h_{2}=s_{2}=3$ and $c_{2}=6$. Therefore, we obtain the
following:

\begin{itemize}
\item[(i)]  The vertical component $X_{2}$ of a Ricci collineation $X\not%
{\equiv }0$ of type 1.1 is a linear combination of the $k_{2}=3$ Killing
vector fields of $\hbox{\ddpp S}^{2}$. On the other hand, the horizontal
component $X_{1}$ satisfies the equation in $k_{R}$ variables (\ref{rr1}).

\item[(ii)]  As the curvature of $\hbox{\ddpp S}^{2}$ is a constant
different from zero, there are not Ricci collineations of types 1.2, 2.2.
Even more, as $s_{2}-h_{2}=0$, there are not Ricci collineations of types
1.3, 2.3 either.

\item[(iii)]  A simple computation shows that $\Delta_{g_{2}}\psi_{2}^{j}=-2%
\psi_{2}^{j}$, $j=4,5,6$. But then, (\ref{r4}) implies $%
(\phi^{2})^{A}_{;A}=R_{2}=2$, which contradicts (\ref{degeneration}).
Therefore, there are not Ricci collineations of type 1.4.

\item[(iv)]  Ricci collineations in Family 2 must satisfy equations (\ref{d4}%
), (\ref{44}), which are in general complicated. If $k_{R}=0$, there are not
Ricci collineations in this family. In fact, in this case (\ref{d4}) (or,
equivalently, (\ref{massutil})) implies again $(\phi^{2})^{A}_{;A}=R_{2}=2$,
in contradiction with (\ref{degeneration}).
\end{itemize}

\section{The degenerate case}

\label{degenerate}

For completeness, we briefly analyse here the cases when Ricci tensor is
degenerate. From (\ref{appear}), the Ricci tensor of a type B warped
spacetime is degenerate if $F\equiv 0$ or $\phi_{A;B}\equiv \frac{\phi}{4}%
R_{1}g_{AB}$ (if both identities hold, the Ricci tensor is zero and any
vector field is a Ricci collineation).

Consider the case $F\equiv 0$ (or, equivalently, $(\phi^{2})^{A}_{;A}\equiv
R_{2}=0$). Then, equations (\ref{lie1})--(\ref{lie3}) show that a vector
field $X=X_{1}+X_{2}\not{\equiv }0$ is a Ricci collineation of the spacetime
if and only if $X_{1}$ is a Killing vector field (perhaps zero) of $%
(M_{1}\times p_{2},g_{R})$ satisfying $R_{AC}X^{C}_{1,\alpha}=0$ for every $%
p_{2}\in M_{2}$. In particular, any Killing vector field of $(M_{1},g_{R})$
is always a Ricci collineation of the spacetime. Moreover, the group of
Ricci collineations becomes infinity, since every vector field $X\not{\equiv
}0 $ with horizontal component $X_{1}\equiv 0$ generates a Ricci
collineation. That is, the vertical component (which is just the component
where Ricci tensor degenerates) of these Ricci collineations is largely
arbitrary (see \cite[Section 2]{camci} for a similar property in
Robertson-Walker spacetimes).

The situation is more complicated when the source of degeneracy is the
identity $\phi_{A;B}\equiv \frac{\phi}{4}R_{1}g_{AB}$. In this case,
Proposition \ref{t2} shows that a vector field $X=X_{1}+X_{2}\not{\equiv }0$
is a Ricci collineation of the spacetime if and only if $X_{2}$ is a
conformal vector field (perhaps zero) of $(p_{1}\times M_{2},g_{2})$
satisfying $R_{\alpha\beta}X^{\beta}_{2,A}=0$ for every $p_{1}\in M_{1}$
and, additionally, $X_{1}$ satisfies (\ref{massutil}) for every $p_{2}\in
M_{2}$. So, in this case we have restrictions on both, the vertical and
horizontal components of $X$. This is due to the dependence of $F$ on both
components, and breaks the similarities with respect to the Robertson-Walker
case.

\section{Conclusion}

\label{sect5}

By analyzing the equations which characterize Ricci collineations of type B
warped spacetimes, we have determined the structure of these symmetries.
They have been classified in eight types according to having or not mixed
variables, and according to their vertical component. As a consequence,
several examples of interest have been considered, and new information about
their collineations has been provided. This study must be understood as an
initial point to begin a systematic computation of Ricci collineations for a
wide family of spacetimes of this class.

As a final remark we would like to point out that it would be very useful to
use computer algebra packages to automate the search of symmetries.
Nevertheless, as far as we know, the available algorithms can only check if
a given vector field is a symmetry or not (at least this is the case of
GRTensor with which we have some familiarity).

\section{Acknowledgments}

We would like to thank Michael Tsamparlis for useful discussions and
comments. We wish to express our gratitude to the referee for the
suggestions which improve the present work. The first author also thanks
CAT-Universidad de Los Andes for the hospitality he received during his stay
there. He was supported by MCyT-FEDER Grant BFM2001-2871-C04-01 and MECyD
Grant EX-2002-0612. The two other authors were supported by
CDCHT-ULA(C-982-99-05-B), CDCHT-ULA (C-1005-00-05-ED) and FONACIT
(S1-2000000820).

\end{document}